\documentclass[twocolumn,nofootinbib,amsmath,amssymb,a4paper]{revtex4}

\usepackage{graphicx}
\usepackage{dcolumn}
\usepackage{bm}

\begin{document}

\title{Quenched lattice calculation of the vector channel $B\rightarrow D^\star \ell \nu$ decay rate}

\author{G.M. de Divitiis$^{a,b}$, R. Petronzio$^{a,b}$, N. Tantalo$^{b,c}$}
\affiliation{\vskip 10pt
$^{a}$~Universit\`a di Roma ``Tor Vergata'', I-00133 Rome, Italy\\
$^{b}$~INFN sezione di Roma ``Tor Vergata'', I-00133 Rome, Italy\\
$^{c}$~Centro Enrico Fermi, I-00184 Rome, Italy
}%

\begin{abstract}
We calculate, in the continuum limit of quenched lattice QCD, the form factor that enters
the decay rate of the semileptonic decay $B\rightarrow D^\star \ell \nu$.  By using
the step scaling method (SSM), previously introduced to handle two scale problems
in lattice QCD, and by adopting flavor twisted boundary conditions we extract 
$F^{B\rightarrow D^\star}(w)$ at finite momentum transfer ($w\ge 1$) and at the physical values 
of the heavy quark masses. Our results can be used in order to extract the CKM matrix element 
$V_{cb}$ by the experimental decay rate without model dependent extrapolations.
The value of $V_{cb}$ agrees with the one obtained from the $B\rightarrow D \ell \nu$ channel
and makes us confident that the quenched approximation well applies to these transitions. 
\end{abstract}

\maketitle

\section{Introduction}
Physics beyond the Standard Model may show up in the hadronic 
flavor sector of the theory and could be revealed by measuring independently the
different entries of the Cabibbo--Kobayashi--Maskawa~\cite{Cabibbo:1963yz,Kobayashi:1973fv} matrix 
and by looking for deviations from unitarity (unitarity triangle analysis, UTA)~\cite{Bona:2008jn}. 
On the theoretical side, a percent relative accuracy on the hadronic matrix elements entering
the different variants of the UTA's~\cite{Bona:2006ah} is needed and lattice 
QCD calculations may eventually provide the required non-perturbative precision.
The study of semileptonic decays of heavy-light mesons mediated by flavor changing $\Delta F=1$ 
hadronic currents gives direct access to CKM matrix elements and is particularly 
convenient from the point of view of lattice QCD calculations. The form factors
entering the decay rates are dimensionless quantities and do not inherit uncertainties 
from the scale setting procedure. Moreover they can be expressed in terms of matrix elements 
undergoing finite and multiplicative renormalization and symmetry arguments can be used to constrain 
numerical results (Ademollo--Gatto theorem, Luke's theorem, spin-symmetry in the heavy quark limit).
Finally, matrix elements involve a single hadron both in initial and final states thus avoiding 
complications due to final state interactions. 

In two previous papers~\cite{de Divitiis:2007ui,de Divitiis:2007uk} we have
calculated, in the quenched approximation of QCD, the form factors entering the decay rate 
of the process $B\rightarrow D\ell\nu$ at non vanishing momentum transfer
and, comparing with experiment, we have extracted the CKM matrix element $V_{cb}$.
In this work we extend our quenched study of heavy-light mesons semileptonic decays by
calculating the form factors entering the decay rate for the process $B\rightarrow D^\star\ell\nu$.
At vanishing momentum transfer the study of pseudoscalar to vector semileptonic transitions
requires the calculation of a single form factor with respect to the two required in the case
of pseudoscalar to pseudoscalar transitions. At finite momentum transfer 
the study of pseudoscalar-vector transitions requires the calculation of a particular
linear combination of four different form factors, $F^{B\rightarrow D^\star}(w)$. 
Here we compute $F^{B\rightarrow D^\star}(w)$ in the interval $1\le w=v_B\cdot v_{D^\star}\le 1.1$
thus allowing the extraction of $V_{cb}$ without extrapolating
experimental data to zero recoil. Although in this case the extrapolation is much less dramatic than
in the pseudoscalar-pseudoscalar channel an important source of systematics can be avoided
by using our results. 

As in refs.~\cite{de Divitiis:2007ui,de Divitiis:2007uk} we make use of the Step Scaling
Method~\cite{Guagnelli:2002jd} devised to reconcile large quark masses with adequate 
lattice resolution and large physical volumes and successfully applied also to the determination of 
heavy quark masses and decay constants~\cite{deDivitiis:2003wy,deDivitiis:2003iy,Guazzini:2007ja}.
We obtain results at non zero momentum transfer with good accuracy by enforcing
special boundary conditions on the quark fields, called flavor twisted~\cite{deDivitiis:2004kq},
that shift by an arbitrary amount the discretized set of lattice momenta 
(see also~\cite{Bedaque:2004kc,Sachrajda:2004mi,Flynn:2005in}).

Our results are not the final ones since they have been obtained within the quenched approximation.
Quenching introduces a systematic error that it is hard to quantify (if not impossible) but allows
us, in view of a future unquenched calculation, to check all the remaining systematics 
(heavy quark methodology, continuum and chiral limits, etc.) and to discuss some technical issues
related to the choice of interpolation operators for vector mesons carrying non vanishing
spatial momenta to be used within the Schr\"odinger Functional formulation of lattice QCD.
Furthermore precise results for $F^{B\rightarrow D^\star}(w)$ at $w>1$ are presently missing even
in the quenched approximation.

To estimate the validity of the quenched approximation we calculate the ratio of physical decay rates 
between vector and pseudoscalar final states. This ratio is indeed a purely QCD observable, independent
from the value of $V_{cb}$, and can be directly compared with experiment.
Our results agree with the measured ratio within experimental errors thus indicating
that residual unquenched corrections are likely within the current experimental uncertainties.

\section{Form factors and decay rate}
The semileptonic decay of a pseudoscalar meson into a vector meson is mediated by
the weak $\mathcal{V}-\mathcal{A}$ current. The relevant matrix elements can be 
parametrized in terms of four form factors. Among possible parameterizations we choose the 
following one
\begin{eqnarray}
&&\frac{\langle \mathcal{M}_V^\alpha \vert \mathcal{V}^\mu \vert \mathcal{M}_P\rangle}{\sqrt{M_V M_P}}
=\varepsilon^{\mu\nu\rho\sigma}\ v_P^\nu v_V^\rho {\epsilon^\star}^\sigma_\alpha\ h_V
\nonumber \\ 
\nonumber \\
\nonumber \\
&&\frac{\langle \mathcal{M}_V^\alpha \vert \mathcal{A}^\mu \vert \mathcal{M}_P\rangle}{\sqrt{M_V M_P}}
=
\nonumber \\
\nonumber \\
&&=i{\epsilon^\star}^\nu_\alpha\ \left[
h_{A_1}(1+w)g^{\mu\nu}-(h_{A_2}v^\mu_P+h_{A_3}v^\mu_V)v^\nu_P
\right] 
\nonumber \\
\nonumber \\
\label{eq:matrixelements}
\end{eqnarray}
where we have used the greek letters $\mu$,$\nu$,$\rho$,$\sigma$ to indicate
Lorentz indices, $M_{P,V}$ are the masses of the pseudoscalar ($\mathcal{M}_P$)
and vector ($\mathcal{M}_V^\alpha$) mesons, $v_{P,V}=p_{P,V}/M_{P,V}$ their $4$-velocities,
$\varepsilon^{\mu\nu\rho\sigma}$ is the totally antisymmetric tensor in four dimensions 
($\varepsilon^{0123}=1$) while $\epsilon^\mu_\alpha$ is the polarization vector of 
$\mathcal{M}_V^\alpha$,
\begin{eqnarray}
&&\sum_{\alpha=1}^3{{\epsilon^\star}^\mu_\alpha \epsilon^\nu_\alpha}=T^{\mu\nu}=-g^{\mu\nu}
+v_V^\mu v_V^\nu
\label{eq:completenesspol}
\end{eqnarray}
The form factors depend upon the masses of the initial and final particles and upon
$w\equiv v_V\cdot v_P$
\begin{eqnarray}
&&h_{V,A_i} \equiv h_{V,A_i}^{P\rightarrow V}(w)\equiv h_{V,A_i}(w,M_P,M_V)
\nonumber \\ \nonumber \\
&&1\le w \le (M_P^2+M_V^2)/2M_PM_V
\end{eqnarray}
In the case where $M_P$ is the $B$ meson mass and $M_V$ is the $D^\star$ meson mass 
the maximum value of $w$ is around $1.5$.

The differential decay rate of the process $B\rightarrow D^\star\ell\nu$, in the case of
massless leptons, is given by
\begin{eqnarray}
&&\frac{d\Gamma^{B\rightarrow D^\star\ell\nu}}{dw}=
\vert V_{cb}\vert^2 \frac{G_F^2}{48\pi^3}M_B^5(1-r)^2r^3 \times
\nonumber \\ \nonumber \\  
&&\qquad\qquad\qquad\sqrt{w^2-1}(1+w)^2
\lambda(w)\left[F^{B\rightarrow D^\star}(w)\right]^2
\nonumber \\
\nonumber \\
\label{eq:decayrate}
\end{eqnarray}
where we have defined $r=M_V/M_P$ and
\begin{eqnarray}
&&t^2(w)=\frac{1-2wr+r^2}{(1-r)^2}
\nonumber \\
\nonumber \\
&&\lambda(w)=1+\frac{4w}{w+1}t^2(w)
\end{eqnarray}
The function $F^{B\rightarrow D^\star}(w)$, 
\begin{eqnarray}
F^{B\rightarrow D^\star}(w)&=&h_{A_1}(w)\sqrt{\frac{H_0^2(w)+H_+^2(w)+H_-^2(w)}{\lambda(w)}}
\nonumber \\
\nonumber \\
\label{eq:fdefinition}
\end{eqnarray}
with
\begin{eqnarray}
X_V(w)&=& \sqrt{\frac{w-1}{w+1}}\frac{h_V(w)}{h_{A_1}(w)}
\nonumber \\ 
\nonumber \\
X_2(w)&=& (w-1)\frac{h_{A_2}(w)}{h_{A_1}(w)}
\nonumber \\ 
\nonumber \\
X_3(w)&=& (w-1)\frac{h_{A_3}(w)}{h_{A_1}(w)}
\nonumber \\ 
\nonumber \\
H_0(w)&=&\frac{w-r-X_3(w)-rX_2(w)}{1-r}
\nonumber \\ 
\nonumber \\
H_\pm(w)&=&t(w)\left[1\mp X_V(w)\right]
\nonumber \\ 
\nonumber \\
\label{eq:fdefinition2}
\end{eqnarray}
is the non perturbative input needed to extract $V_{cb}$ by the measurement of the decay rate.

By noting that at zero recoil $F^{B\rightarrow D^\star}(1)$ is identically equal to $h_{A_1}(1)$
the complexity of the theoretical calculation can be considerably reduced since, in this 
particular case, a single matrix element with initial and final particles both at rest is needed
instead of the four matrix elements at non vanishing momentum transfer required to solve 
the full system of eqs.~(\ref{eq:matrixelements}) with respect to $h_{V,A_i}$. 
For these reasons previous lattice studies have been devoted to the calculation
of $F^{B\rightarrow D^\star}(w)$ at zero recoil only, where it can be extracted with good
statistical accuracy both in the quenched approximation~\cite{Hashimoto:2001nb} and in 
the $n_f=2+1$ unquenched theory~\cite{Laiho:2007pn}. 
On the other hand, it is not possible to measure directly the
decay rate at zero recoil because of the presence of the kinematical factor $(w-1)^{1/2}$ in 
eq.~(\ref{eq:decayrate}) and experimental data at $w=1$ are obtained by extrapolation. The systematics
introduced by this extrapolation is much less dramatic with respect to the case of the
decay $B\rightarrow D\ell\nu$ where the kinematical suppression goes like $(w-1)^{3/2}$ but can
be nevertheless eliminated. In the following we calculate $F^{B\rightarrow D^\star}(w)$
in the range $1\le w \le 1.1$ that includes values of $w$ where experimental data are directly
available with good precision.

\section{Schr\"odinger Functional correlators} \label{sec:notations}
We have carried out the calculation within the $O(a)$ improved Schr\"odinger Functional 
formalism~\cite{Luscher:1992an,Sint:1993un} with vanishing background fields. 
The choice of the Schr\"odinger Functional regularization is particularly convenient to perform
simulations on small physical volumes (see section~\ref{sec:ssm})
because Dirichelet boundary conditions in the time direction
provide an infrared regulator that allows the simulation of massless quarks. At the same time,
the extraction of physical matrix elements involving vector mesons at non vanishing
spatial momenta from Schr\"odinger Functional correlators requires some additional care 
with respect to the case of quark fields satisfying periodic boundary conditions in the time 
direction.

\subsection{Boundary and bulk operators}
In defining interpolation operators of meson states we need to
distinguish between bulk fields $\psi(x)$, boundary fields $\zeta(x)$ living at $x_0=0$ 
\begin{eqnarray}
P_+\ \zeta(\vec{x}) &=& \frac{1+\gamma_0}{2}\ \psi(0,\vec{x}) =0
\nonumber \\ \nonumber \\
\bar{\zeta}(\vec{x})\ P_- &=& \bar{\psi}(0,\vec{x})\ \frac{1-\gamma_0}{2} = 0
\end{eqnarray}
and boundary fields $\zeta^\prime(x)$ living a t $x_0=T$
\begin{eqnarray}
P_-\ \zeta^\prime(\vec{x}) &=& \frac{1-\gamma_0}{2}\ \psi(T,\vec{x}) =0
\nonumber \\ \nonumber \\
\bar{\zeta}^\prime(\vec{x})\ P_+ &=& \bar{\psi}(T,\vec{x})\ \frac{1+\gamma_0}{2} = 0
\end{eqnarray}
Different quark flavors will be  distinguished, if needed, by using explicit indexes.
External momenta have been set by using flavor twisted b.c. for the 
heavy flavors. In particular we have used
\begin{eqnarray}
&&\psi(x+\hat{1}L)=e^{i\theta}\psi(x)
\nonumber \\ 
\nonumber \\
&&p_1=\frac{\theta}{L}+\frac{2\pi k_1}{L},\qquad k_1\in \mathbb{N}
\end{eqnarray}
with different values of $\theta$ for the different heavy quarks
and ordinary periodic b.c. in the other spatial directions and for the light quarks.

A generic meson state on the boundaries can be expressed as a bilinear field operator acting on the vacuum. 
In the case of vector mesons, the Schr\"odinger Functional boundary conditions select a particular
combination between the two possible choices  $V^\mu(x)=\bar{\psi}(x)\gamma^\mu\psi(x)$ and
$T^{0\mu}(x)=\bar{\psi}(x)\gamma^0\gamma^\mu\psi(x)$. On the wall at $x_0=0$ we have
\begin{eqnarray}
\mathbf{V}^\mu&=&\frac{a^6}{L^3}\sum_{{\bf y},{\bf z}}{\bar{\zeta}({\bf y})\gamma^\mu\zeta({\bf z})}
\nonumber \\ 
\nonumber \\
&=&\frac{a^6}{L^3}\sum_{{\bf y},{\bf z}}{\bar{\psi}(0,{\bf y})\frac{
2\gamma^\mu
-2\gamma^0g^{\mu0}+[\gamma^0,\gamma^\mu]
}{4}\psi(0,{\bf z})}
\nonumber \\
\label{eq:vectorinterp}
\end{eqnarray}
Indeed the two boundary projectors act on $\gamma^\mu$ by killing its time component
and by introducing a mixing with the tensor $\sigma^{0\mu}=i[\gamma^0,\gamma^\mu]/2$,
\begin{equation}
P_+\gamma^\mu P_- =
\left\{
\begin{array}{ll}
0, & \mu=0 \\
\\
\frac{1}{2} (\gamma^i - i \sigma^{0i}), 
& \mu=i=1,2,3 \\
\end{array}
\right.
\end{equation}
which, in compact notation, is equivalent to the gamma matrix combination appearing
in eq.~(\ref{eq:vectorinterp})
\begin{eqnarray}
&& P_+\gamma^\mu P_-  =
\frac{2\gamma^\mu-2\gamma^0g^{\mu0}+[\gamma^0,\gamma^\mu]}{4}   
\end{eqnarray}
The matrix elements of $\mathbf{V}^\mu$ between the vacuum and a vector meson state,
entering the spectral decomposition of two and three point correlation functions,
can be thus parametrized as
\begin{eqnarray}
\langle 0 \vert \mathbf{V}^\mu \vert \mathcal{M}_V^\alpha \rangle
= \rho_V\left[\epsilon^\mu_\alpha-\epsilon^0_\alpha g^{0\mu}\right]
+\rho_T\left[\epsilon^\mu_\alpha v^0_V-\epsilon^0_\alpha v^\mu_V\right]
\nonumber \\
\end{eqnarray}
The projection of the wall sources on the physical states entail unknown non-perturbative
wave functions, $\rho_V$ and $\rho_T$, that cancel out exactly in our choice of ratios
of correlation functions discussed in the next subsection.
Pseudoscalar mesons do not carry polarization indexes and the effect of the boundary conditions
can be reabsorbed into a redefinition of the wave function,
\begin{eqnarray}
\mathbf{P}&=&\frac{a^6}{L^3}\sum_{{\bf y},{\bf z}}{\bar{\zeta}({\bf y})\gamma_5\zeta({\bf z})}
\nonumber \\ 
\nonumber \\
&=&\frac{a^6}{L^3}\sum_{{\bf y},{\bf z}}{
\bar{\psi}(0,{\bf y})\frac{\gamma_5+\gamma_0\gamma_5}{2}\psi(0,{\bf z})}
\nonumber \\
\nonumber \\
\nonumber \\
\langle 0 \vert \mathbf{P} \vert \mathcal{M}_P \rangle
&=&\rho_P + \rho_A\ v_P^0
\nonumber \\
\nonumber \\
&=&\tilde{\rho}_P
\end{eqnarray}
The relations above hold at  $x_0=0$ but
the same arguments can be repeated for the boundary operators on the wall at $x_0=T$ that in the following 
will be called $\mathbf{P^\prime}$ and $\mathbf{V^\prime}^\mu$.

For later use, we also define the improved bulk operators
\begin{eqnarray}
&&\mathcal{A}^\mu(x_0)=A^\mu(x)+ac_A\frac{\partial_\mu+\partial_\mu^*}{2}P(x)
\nonumber \\ \nonumber \\
&&\mathcal{V}^\mu(x)=V^\mu(x)+ac_V\frac{\partial_\nu+\partial_\nu^*}{2}T^{\mu\nu}(x)
\end{eqnarray}
where $P(x)=\bar{\psi}(x)\gamma_5\psi(x)$ is the pseudoscalar density, 
$A^\mu(x)=\bar{\psi}(x)\gamma^\mu\gamma_5\psi(x)$ the axial current and
the improvement coefficients $c_A$ and $c_V$ have been taken from 
refs.~\cite{Luscher:1996ug,Guagnelli:1997db,Sint:1997jx}.

\subsection{Quark masses}
Particle states are fixed by tuning the values of the
corresponding quark masses. We use three definitions of renormalization
group invariant quark masses differing at finite lattice spacing by terms of order $O(a^2)$.
We have calculated the following two point correlation functions
\begin{eqnarray}
f^A_{rr}(x_0)&=&\sum_{\bf x}{\langle \mathbf{P}_{rr} A^0_{rr}(x)\rangle}
\nonumber \\
\nonumber \\
f^P_{rr}(x_0)&=&-\sum_{\bf x}{\langle \mathbf{P}_{rr} P_{rr}(x)\rangle}
\end{eqnarray}
and defined
\begin{eqnarray}
am_r^{AWI}&=&\frac{1}{2f^P_{rr}}\left[\frac{\partial_0+\partial_0^*}{2}f^A_{rr}+ac_A\partial_0\partial_0^*f^P_{rr}\right]
\label{eq:pcac} \\
\nonumber \\
am_r^{b}&=&\frac{1}{2}\left[\frac{1}{k_r}-\frac{1}{k_c}\right]
\end{eqnarray}
where $a$ is the lattice spacing, $k_r$ is the hopping parameter of the $r$ quark and
$k_c$ is the critical value of the hopping parameter.

A first definition of renormalization group invariant (RGI) quark masses 
has been obtained by the following relation
\begin{eqnarray}
m_r = Z_M\;\left[1+(b_A-b_P)\ am_r^{b} \right]\; m^{AWI}_r
\label{eq:rgi1}
\end{eqnarray}
The combination $b_A-b_P$ of the improvement coefficients of the axial current and pseudoscalar density has been computed non-perturbatively 
in~\cite{deDivitiis:1997ka,Guagnelli:2000}. The factor $Z_M$ 
is known with very high precision in a range of inverse bare couplings that does not cover all 
the values of $\beta$ used in our simulations.
We have used the results reported in table~6 of ref.~\cite{Capitani:1998mq} 
to parametrize $Z_M$ in the enlarged range of $\beta$ values $[5.9,7.6]$.
A second definition of RGI quark mass has been obtained by the relation
\begin{eqnarray}
m_r = Z_M\;Z\;\left[1+b_m\ am_r^{b} \right]\; m_r^{b}
\label{eq:rgi2}
\end{eqnarray}
where the improvement coefficient $b_m$ and the renormalization constant $Z$ has been
also taken from refs.~\cite{deDivitiis:1997ka,Guagnelli:2000}.
An additional definition of RGI quark masses has been obtained by using improved
lattice derivatives in eq.~(\ref{eq:pcac}).

\subsection{Three point correlators}
The matrix elements in eqs.~(\ref{eq:matrixelements}) can be calculated on
the lattice by building suitable ratios of the following three point correlation functions
\begin{eqnarray}
\langle PVP\rangle^\mu_{if}(x_0,\vec{p}_{P_i},\vec{p}_{P_f})&=&
Z_V^I\sum_{\vec{x}}{\langle
\mathbf{P}_{li}\ \mathcal{V}^\mu_{if}(x) \ \mathbf{P^\prime}_{fl}
\rangle}
\nonumber \\
\nonumber \\
\langle VVV\rangle^{I\mu I}_{if}(x_0,\vec{p}_{V_i},\vec{p}_{V_f})&=&
Z_V^I\sum_{\vec{x}}{\langle
\mathbf{V}_{li}^I\ \mathcal{V}^\mu_{if}(x) \ \mathbf{V^\prime}_{fl}^I
\rangle}
\nonumber \\
\nonumber \\
\langle PVV\rangle^{\mu I}_{if}(x_0,\vec{p}_{P_i},\vec{p}_{V_f})&=&
Z_V^I\sum_{\vec{x}}{\langle
\mathbf{P}_{li}\ \mathcal{V}^\mu_{if}(x) \ \mathbf{V^\prime}_{fl}^I
\rangle}
\nonumber \\
\nonumber \\
\langle PAV\rangle^{\mu I}_{if}(x_0,\vec{p}_{P_i},\vec{p}_{V_f})&=&
Z_A^I\sum_{\vec{x}}{\langle
\mathbf{P}_{li}\ \mathcal{A}^\mu_{if}(x) \ \mathbf{V^\prime}_{fl}^I
\rangle}
\nonumber \\
\end{eqnarray}
In the previous equations we have explicitly indicated heavy ($i,f$) and
light ($l$) flavor indices and we have implicitly defined
\begin{eqnarray}
Z_V^I&=&Z_V\left(1+b_V\frac{am_i^{b}+am_f^{b}}{2}\right)
\nonumber \\
\nonumber \\
Z_A^I&=&Z_A\left(1+b_A\frac{am_i^{b}+am_f^{b}}{2}\right)
\end{eqnarray}
Let us now consider the normalization factors 
\begin{eqnarray}
&&N^I_{if}(x_0,\vec{p}_{P_i},\vec{p}_{V_f})
\nonumber \\
\nonumber \\
&&=\sqrt{\langle PVP\rangle^0_{ii}(x_0,\vec{p}_{P_i},\vec{p}_{P_i})
\langle VVV\rangle^{I0I}_{ff}(x_0,\vec{p}_{V_f},\vec{p}_{V_f})}
\nonumber \\
\end{eqnarray}
Because of our choice of spatial momenta (having non vanishing components
along the direction $\hat{1}$ only) the normalization factors $N^1$ will in general be
different from the remaining ones, $N^{2,3}$. By assuming single state dominance, by relying
on the conservation of the vector current and by using the form factors definition, 
eqs.~(\ref{eq:matrixelements}), and the completeness relation of the polarization 
vectors, eq.~(\ref{eq:completenesspol}), we have
\begin{eqnarray}
&&\mathcal{C}^\perp =
\sqrt{v^0_Pv^0_V}\left\{
\frac{\langle PAV\rangle^{22}(T/2,\vec{p}_P,\vec{p}_V)}{N^2(T/2,\vec{p}_P,\vec{p}_V)}
\right.
\nonumber \\
\nonumber \\
&&
\left. \qquad\qquad\qquad\quad+
\frac{\langle PAV\rangle^{33}(T/2,\vec{p}_P,\vec{p}_V)}{N^3(T/2,\vec{p}_P,\vec{p}_V)}
\right\}
\nonumber \\
\nonumber \\
&&=(1+w)h_{A_1}
\nonumber \\ 
\nonumber \\
\nonumber \\
&&\mathcal{B}^\perp =
\sqrt{v^0_Pv^0_V}\left\{
\frac{\langle PVV\rangle^{32}(T/2,\vec{p}_P,\vec{p}_V)}{N^2(T/2,\vec{p}_P,\vec{p}_V)}
\right.
\nonumber \\
\nonumber \\
&&
\left. \qquad\qquad\qquad\quad-
\frac{\langle PVV\rangle^{23}(T/2,\vec{p}_P,\vec{p}_V)}{N^3(T/2,\vec{p}_P,\vec{p}_V)}
\right\}
\nonumber \\
\nonumber \\
&&=\sqrt{w^2-1}h_{V}
\end{eqnarray}
The extraction of $h_{A_1}$ from the first of the previous relations is straightforward.
In view of the calculation of the decay rate (see eqs.~(\ref{eq:decayrate}), 
(\ref{eq:fdefinition}) and~(\ref{eq:fdefinition2})),
we do not remove the factor $\sqrt{w^2-1}$ from the second equation above. More precisely,
we measure on the lattice the following combinations
\begin{eqnarray} 
&&h_{A_1}=\frac{\mathcal{C}^\perp}{1+w}
\label{eq:derivation1}
\\
\nonumber \\
\nonumber \\
&&X_V=\sqrt{\frac{w-1}{w+1}}\frac{h_V}{h_{A_1}}=\frac{\mathcal{B}^\perp}{\mathcal{C}^\perp}
\end{eqnarray} 
that enter directly the definition of $F^{B\rightarrow D^\star}(w)$.
In the case of the remaining form factors, $h_{A_2}$ and $h_{A_3}$, we follow a similar
path. By defining
\begin{eqnarray}
\mathcal{C}^0&=& 2\sqrt{v^0_Pv^0_V}\
\frac{\langle PAV\rangle^{01}(T/2,\vec{p}_P,\vec{p}_V)}{N^1(T/2,\vec{p}_P,\vec{p}_V)}
\nonumber \\
\nonumber \\
&=&(1+w)h_{A_1}v^1_V+\sqrt{w^2-1}\left[h_{A_2}v^0_P+h_{A_3}v^0_V \right]
\nonumber \\ 
\nonumber \\
\nonumber \\
\mathcal{C}^1&=& 2\sqrt{v^0_Pv^0_V}\
\frac{\langle PAV\rangle^{11}(T/2,\vec{p}_P,\vec{p}_V)}{N^1(T/2,\vec{p}_P,\vec{p}_V)}
\nonumber \\
\nonumber \\
&=&(1+w)h_{A_1}v^0_V+\sqrt{w^2-1}\left[h_{A_2}v^1_P+h_{A_3}v^1_V \right]
\nonumber \\
\nonumber \\
\end{eqnarray}
we measure on the lattice the following combinations of
correlation functions
\begin{eqnarray}
&&X_2=(w-1)\frac{h_{A_2}}{h_{A_1}}=\frac{v^0_V\mathcal{C}^1-v^1_V\mathcal{C}^0}{\mathcal{C}^\perp}-1
\label{eq:derivation2}
\\
\nonumber \\
\nonumber \\
&&X_3=(w-1)\frac{h_{A_3}}{h_{A_1}}=\frac{v^1_P\mathcal{C}^0-v^0_P\mathcal{C}^1}{\mathcal{C}^\perp}+w
\label{eq:derivation3}
\end{eqnarray}
that provide the combinations of form factors entering the definition of $F^{B\rightarrow D^\star}(w)$.

A key point in reducing statistical fluctuations in the extraction of the different combinations
of form factors is the definition of the $4$-velocities, and consequently of $w=v_P\cdot v_V$, 
entering eqs.~(\ref{eq:derivation1}), (\ref{eq:derivation2}) and~(\ref{eq:derivation3}). 
A particularly convenient choice is the definition of velocities in terms of three point 
correlation functions by the following ratios
\begin{eqnarray}
\frac{p_{V_i}}{E_{V_i}}&=& \frac{1}{3}
\frac{\langle VVV\rangle^{111}_{ii}(T/2,\vec{p}_{V_i},\vec{p}_{V_i})}
{\langle VVV\rangle^{101}_{ii}(T/2,\vec{p}_{V_i},\vec{p}_{V_i})}
\nonumber \\
\nonumber \\
&+&\frac{1}{3}
\frac{\langle VVV\rangle^{212}_{ii}(T/2,\vec{p}_{V_i},\vec{p}_{V_i})}
{\langle VVV\rangle^{202}_{ii}(T/2,\vec{p}_{V_i},\vec{p}_{V_i})}
\nonumber \\
\nonumber \\
&+&\frac{1}{3}
\frac{\langle VVV\rangle^{313}_{ii}(T/2,\vec{p}_{V_i},\vec{p}_{V_i})}
{\langle VVV\rangle^{303}_{ii}(T/2,\vec{p}_{V_i},\vec{p}_{V_i})}
\\
\nonumber \\
\nonumber \\
\frac{p_{P_i}}{E_{P_i}}&=& 
\frac{\langle PVP\rangle^{1}_{ii}(T/2,\vec{p}_{P_i},\vec{p}_{P_i})}
{\langle PVP\rangle^{0}_{ii}(T/2,\vec{p}_{P_i},\vec{p}_{P_i})}
\end{eqnarray}
This way we have been able to define all the form factors
entirely in terms of three point correlation functions and to keep under control
statistical fluctuations also at non vanishing momentum transfer.

\section{the step scaling method}\label{sec:ssm}

The idea behind the SSM is to split the complexity of the calculation of quantities
depending upon two largely separated energy scales into several calculations
performed on different physical volumes. The small volume calculations are needed in order to
resolve the dynamics of the heavy quarks without recurring to any approximation but
introducing, at intermediate stages, finite volume effects (FVE)
\begin{eqnarray}
\mathcal{O}(physical)\; =\; \mathcal{O}(finite\ volume)\;\times\; FVE 
\end{eqnarray}
The finite volume effects are subsequently accounted for
by performing simulations on progressively larger volumes. The success of this strategy
depends on the details of the problem and hence on the possibility of
computing the finite volume observable, finite volume effects and their
product with smaller errors and systematics with respect to the
ones that would be obtained by a direct calculation. The strength of the
method is a great freedom in the definition of the observable on finite volumes
provided that its physical value is recovered at the end of the procedure.

In the calculation of heavy-light meson observables the two scales are the mass
of the heavy quarks ($b$,$c$) and the mass of the light quarks ($u$,$d$,$s$). 
Here we consider the form factor $F^{P\rightarrow V}(w)$ as a 
function of $w$, the volume, $L^3\times T$, and identify heavy meson states by the corresponding RGI 
quark masses that in the infinite volume limit lead to the physical meson 
spectrum~\cite{deDivitiis:2003iy}. 

First we compute the observable $F^{B\rightarrow D^\star}(w;L_0^3\times L_0)$ on a small volume
$L_0^3\times L_0$, $T=L$, $L_0\simeq 0.4$~fm, which is chosen to accommodate the 
dynamics of the $b$-quark.
The observable on the finite volume is defined as explained in the previous section, i.e.
by relying on single state dominance of the different correlation functions. Of course
this is not true at $x_0=L_0/2\simeq 0.2$~fm but, letting $x_0$ to scale proportionally
to the volume in the finite size scaling iteration, becomes true by removing finite volume
effects.

A first portion of finite volume effects is removed by evolving the volume from $L_0^3\times L_0$ to 
$L_1^3\times L_1$, $L_1=2L_0$, by the ratio
\begin{eqnarray} 
\sigma^{P\rightarrow D^\star}(w;L_0,L_1)=\frac{F^{P\rightarrow D^\star}(w;L_1^3\times L_1)}
{F^{P\rightarrow D^\star}(w;L_0^3\times L_0)}
\label{eq:sigma1}
\end{eqnarray}
The crucial point is that the step scaling functions are calculated by simulating
heavy quark masses $m_P$ smaller than the $b$-quark mass. The physical
value $\sigma^{B\rightarrow D^\star}(w;L_0,L_1)$ is obtained by a smooth extrapolation in $1/m_P$ that
relies on the HQET expectations and upon the general idea that finite volume effects, measured
by the $\sigma$'s, are almost insensitive to the high energy scale. 

The final result is obtained by further evolving the volume from $L_1^3\times L_1$ 
to $L_2^3\times T_2$, with $L_2=4L_0$ and $T_2=3L_2/2$ by calculating a second step 
scaling function
\begin{eqnarray} 
\sigma^{P\rightarrow D^\star}(w;L_1,L_2)=\frac{F^{P\rightarrow D^\star}(w;L_2^3\times 3L_2/2)}
{F^{P\rightarrow D^\star}(w;L_1^3\times L_1)}
\nonumber \\
\label{eq:sigma2}
\end{eqnarray}
and by the following identity
\begin{eqnarray} 
F^{B\rightarrow D^\star}(w;L_2^3\times 3L_2/2)&=&
F^{B\rightarrow D^\star}(w;L_0^3\times L_0)
\nonumber \\
\nonumber \\
&\times&\sigma^{B\rightarrow D^\star}(w;L_0,L_1)
\nonumber \\ 
\nonumber \\
&\times&\sigma^{B\rightarrow D^\star}(w;L_1,L_2)
\label{eq:infresultformula}
\end{eqnarray}
%

\section{lattice simulations}\label{sec:simulations}

\begin{table}[h]
\begin{ruledtabular}
\begin{tabular}{cccc}
 & $\beta$ & $L^3\times T$ & $N_{cnfg}$\\
\hline
$L_0A$ & 7.6547 & $32^3\times 32$   & 433  \\
$L_0B$ & 7.4082 & $24^3\times 24$   & 298  \\
$L_0C$ & 7.2611 & $20^3\times 20$   & 128  \\
\hline
$L_0a$ & 7.0203 & $16^3\times 16$   & 293  \\
$L_0b$ & 6.7750 & $12^3\times 12$   & 640  \\
$L_0c$ & 6.4956 & $ 8^3\times  8$   & 1600 \\[1ex]

$L_1A$ & 7.0203 & $32^3\times 32$   & 326  \\
$L_1B$ & 6.7750 & $24^3\times 24$   & 243  \\
$L_1C$ & 6.4956 & $16^3\times 16$   & 120  \\
\hline
$L_1a$ & 6.4956 & $16^3\times 16$   & 1051 \\
$L_1b$ & 6.2885 & $12^3\times 12$   & 3168 \\
$L_1b$ & 6.0219 & $ 8^3\times  8$   & 720  \\[1ex]

$L_2A$ & 6.4956 & $32^3\times 48$   & 152\\
$L_2B$ & 6.2885 & $24^3\times 36$   & 150\\
$L_2C$ & 6.0219 & $16^3\times 24$   & 200\\
\end{tabular}
\end{ruledtabular}
\caption{\label{tab:sims}
Table of lattice simulations.}
\end{table}

\subsection{Small volume}

\begin{figure}[h]
\includegraphics[width=0.45\textwidth]{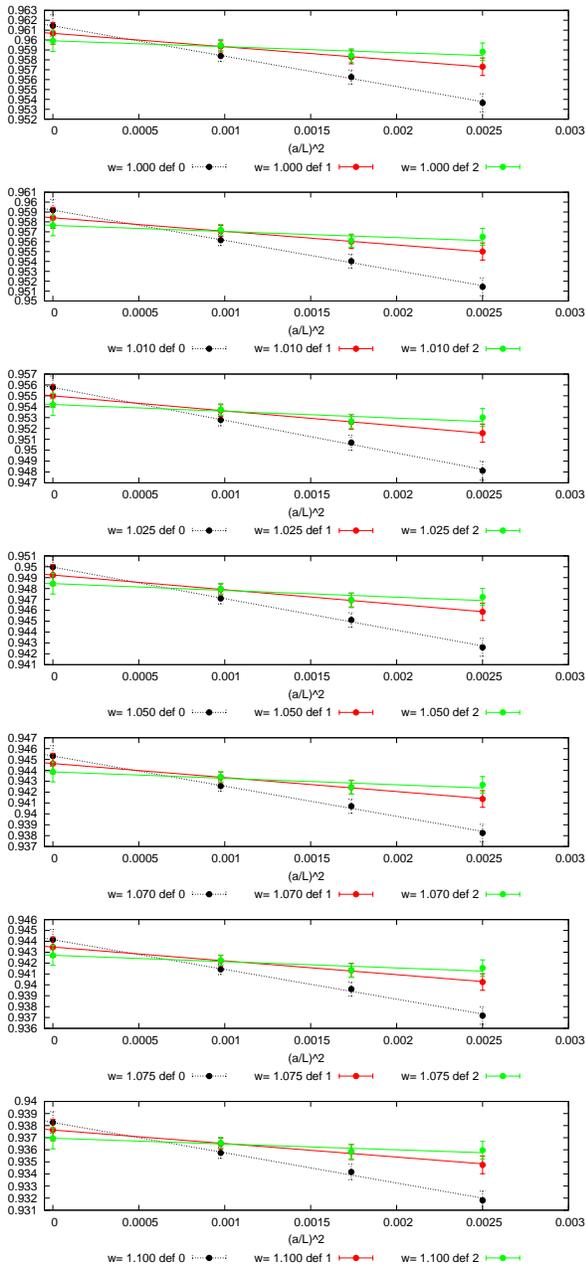}
\caption{\label{fig:continuums0} 
Continuum extrapolations of $F^{B\rightarrow D^\star}(w;L_0^3\times L_0)$
for the different simulated values of $w$. The different fits are linear in
$(a/L_0)^2$ and correspond to different definitions of RGI quark masses.}
\end{figure}
\begin{figure}[h]
\includegraphics[width=0.45\textwidth]{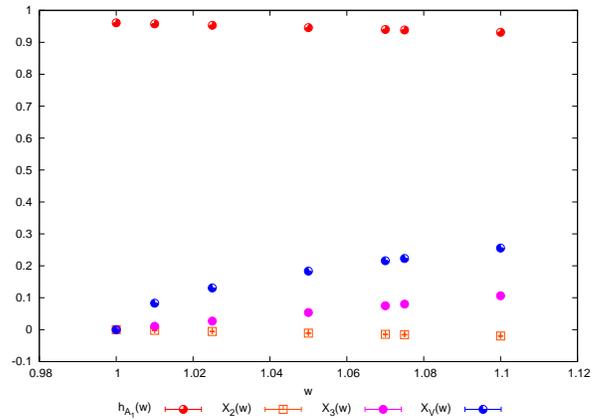}
\caption{\label{fig:ffss0} 
Small volume results for $h_{A_1}(w)$, $X_V(w)$,
$X_2(w)$ and $X_3(w)$. On this scale the errors are much smaller than the symbols.}
\end{figure}

Simulations on the small volume $L_0^3\times L_0$ have been carried out at three different
values of the lattice spacing corresponding to the entries $L_0A$, $L_0B$ and $L_0C$
in TABLE~\ref{tab:sims}. The beta values have been deduced from 
refs~\cite{Della Morte:2006cb,Della Morte:2005yc,Heitger:2003ue} and have been matched
by fixing the value of the renormalized strong coupling constant in the Schr\"odinger
Functional scheme. The physical extension of the volume is $L_0\simeq 0.718r_0$ where
$r_0\simeq 0.5$~fm~\cite{Sommer:1993ce}.

On such a small volume we have simulated massless light quarks and we do not need to discuss
the systematics associated to chiral extrapolations. 

In FIG.~\ref{fig:continuums0} we show the continuum extrapolation of 
$F^{B\rightarrow D^\star}(w;L_0^3\times L_0)$ for the seven different values of $w$ that
have been simulated. Different fits correspond to the three different definitions of quark masses
introduced in section~\ref{sec:notations}: the extrapolations must coincide within the errors
as happens to be in practice and the continuum results are defined by a jackknife average.
In general we combine results of different simulations in big jackknife samples according
to the recipe discussed in the Appendix A of ref.~\cite{DelDebbio:2007pz}.

In FIG.~\ref{fig:ffss0} we show the different form factors that enter into the definition of
$F^{B\rightarrow D^\star}(w;L_0^3\times L_0)$ already extrapolated to the continuum limit.
As expected, $X_V(w)$, $X_2(w)$ and $X_3(w)$ vanish at zero recoil while $h_{A_1}(w)$ is of
order one. The scale of the figure does not allow to distinguish the errors that are much
smaller than the symbols but it allows to appreciate the relative sizes of the different
form factors.

Concerning the renormalization factors, our definitions of $h_{A_1}(w)$ and $X_V(w)$ require the
knowledge of $Z_A/Z_V$ and of $(b_A-b_V)$ while, in the case of $X_2(w)$ and $X_3(w)$, these cancel
out in the ratios. Unfortunately the ratio $Z_A/Z_V$ has not been computed directly and we have used the
separate non-perturbative determinations of $Z_A$ and $Z_V$ performed in ref.~\cite{Luscher:1996jn}.
Also $b_V$ has been non-perturbatively determined in ref.~\cite{Luscher:1996jn} while for
$b_A$ we use its perturbative value (approximation well justified at this small values
of bare couplings and, a posteriori, by looking at the continuum extrapolations of 
FIG.~\ref{fig:continuums0}). The overall systematics due to renormalization factors 
is largely accounted for by adding a $0.6\%$ relative error to our final result
(see ref.~\cite{Luscher:1996jn}).
Since renormalization factors cancel out in the definition of the step scaling 
functions there are no other systematics associated to the renormalization procedure in our 
calculation.

\subsection{Intermediate volume}
The step toward the intermediate volume $L_1^3\times L_1$, $L_1=2L_0$, has been performed
by calculating the denominator of eq.~(\ref{eq:sigma1}) at three different lattice spacings,
corresponding to the entries $L_0a$, $L_0b$ and $L_0c$ of TABLE~\ref{tab:sims}, and
the numerator with the same lattice spacings but with twice the number of lattice points
per direction, entries $L_1A$, $L_1B$ and $L_1C$ of TABLE~\ref{tab:sims}. Also for this step
the light quark has been simulated at zero mass. The value of the heaviest quark mass simulated
on these volumes has been halved with respect to the previous step in order to have the
same order of cutoff effects.

\begin{figure}[h]
\includegraphics[width=0.45\textwidth]{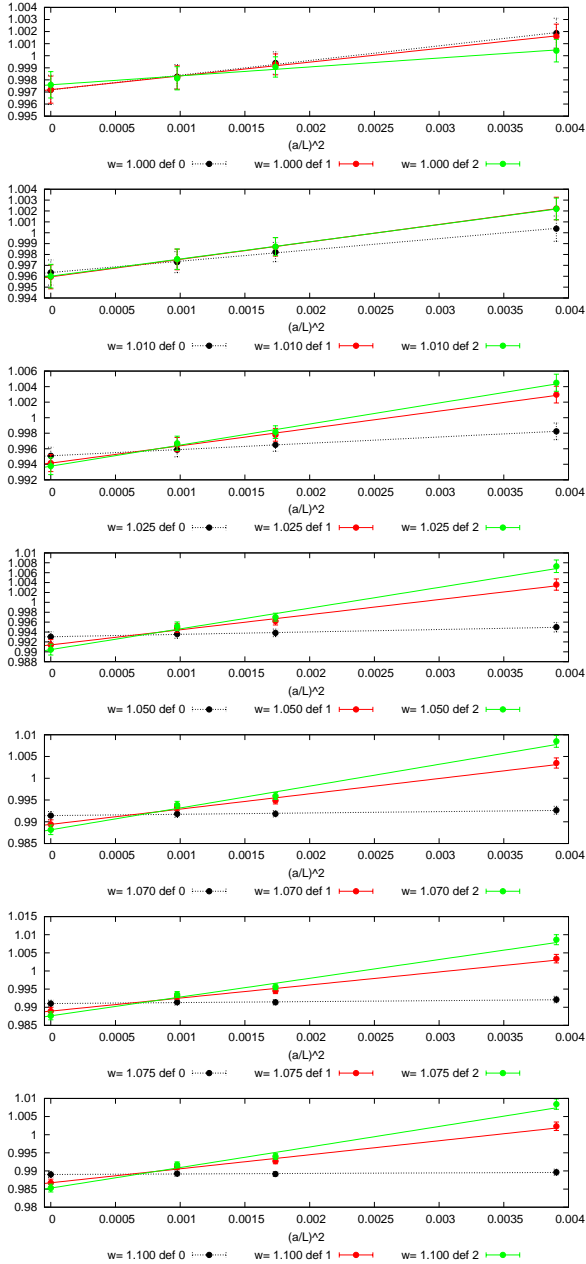}
\caption{\label{fig:continuums1} 
Continuum extrapolations of $\sigma^{P\rightarrow D^\star}(w;L_0,L_1)$ for
the heaviest quark mass ($m_P\simeq m_b/2$) simulated on this step and
for the different simulated values of $w$. The different fits are linear in
$(a/L_0)^2$ and correspond to different definitions of RGI quark masses.}
\end{figure}
In FIG.~\ref{fig:continuums1} we show the continuum extrapolation of 
$\sigma^{P\rightarrow D^\star}(w;L_0,L_1)$ for the seven different values of $w$ that
have been simulated. The pseudoscalar meson state correspond to a quark having a mass of 
about half of the physical value of the $b$-quark mass ($m_P\simeq m_b/2$) i.e. 
the heaviest mass simulated on these volumes; 
similar plots could have been shown for the other simulated heavy quark masses. 
Different fits correspond to the three different definitions of quark masses
introduced in section~\ref{sec:notations}: the continuum results are defined by the 
jackknife average of the independent extrapolations.

\begin{figure}[t]
\includegraphics[width=0.45\textwidth]{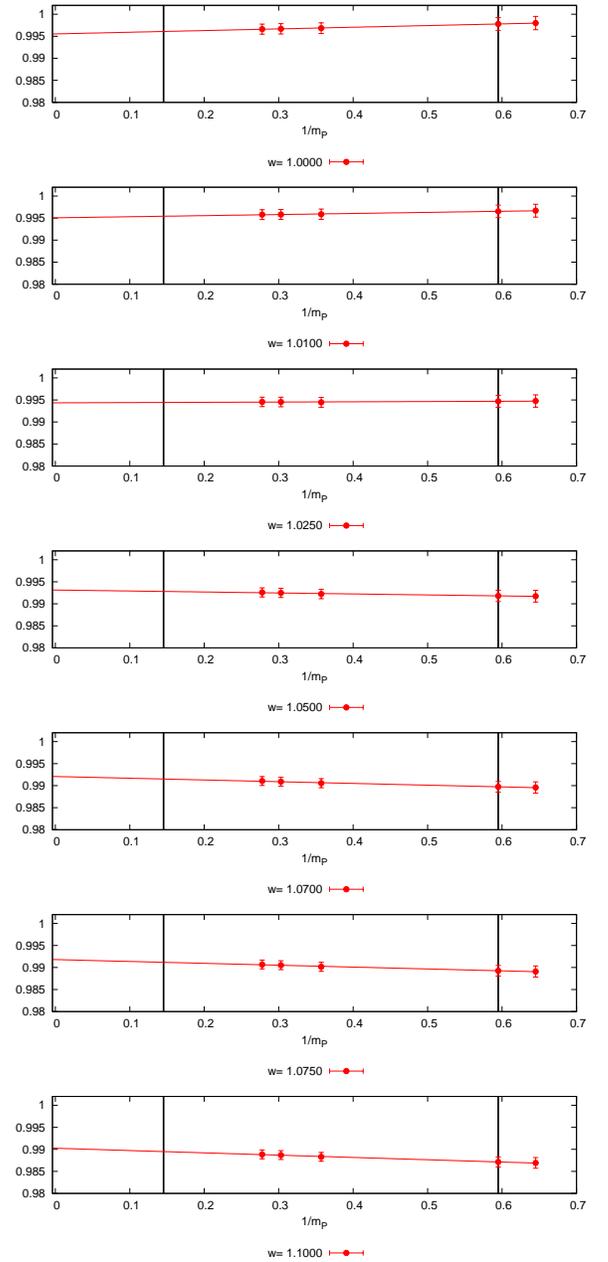}
\caption{\label{fig:references1} 
Continuum step scaling function $\sigma^{P\rightarrow D^\star}(w;L_0,L_1)$ as a function
of the inverse RGI heavy quark mass of the pseudoscalar state, $1/m_P$,
for the different values of $w$. The vertical black lines represent the physical values
of the charm and bottom quark masses.}
\end{figure}
In FIG.~\ref{fig:references1} we can test our hypothesis of the low sensitivity of the
finite volume effects upon the heavy quark mass. The figure shows the step scaling
function $\sigma^{P\rightarrow D^\star}(w;L_0,L_1)$ in the continuum limit and at fixed $w$
as a function of the inverse RGI heavy quark mass of the pseudoscalar state $1/m_P$.
The RGI heavy quark mass of the vector state has been held fixed to its physical value $m_c$.
The physical step scaling functions are obtained by linear extrapolations so mild that the
values at $m_b$ differ by the simulated ones by a few per mille.

\subsection{Final volume}
\begin{figure}[t]
\includegraphics[width=0.45\textwidth]{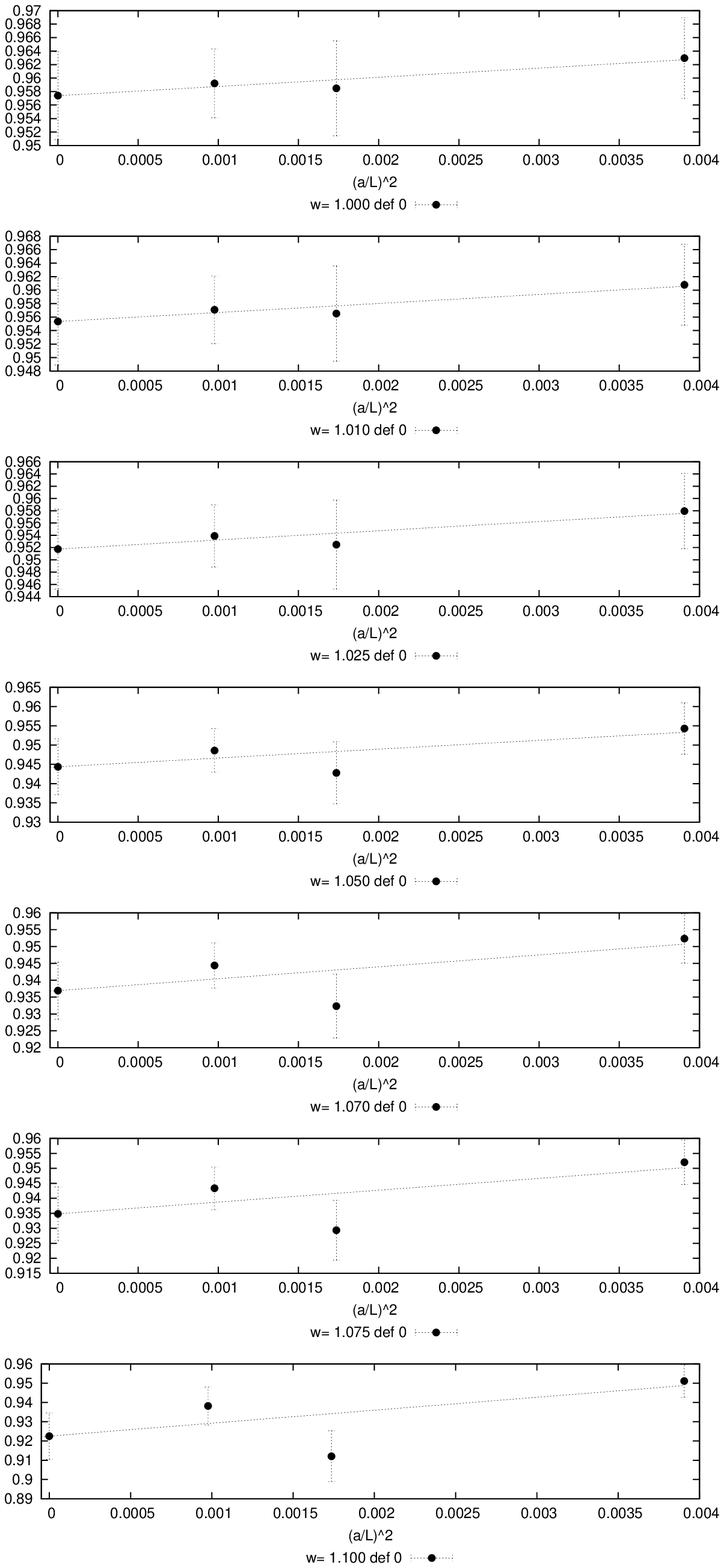}
\caption{\label{fig:continuums2} 
Continuum extrapolations of $\sigma^{P\rightarrow D^\star}(w;L_1,L_2)$ for
the heaviest quark mass ($m_P\simeq m_b/4$) simulated on this step, $m_l=m_s$ and
for the different simulated values of $w$.}
\end{figure}
\begin{figure}[h]
\includegraphics[width=0.45\textwidth]{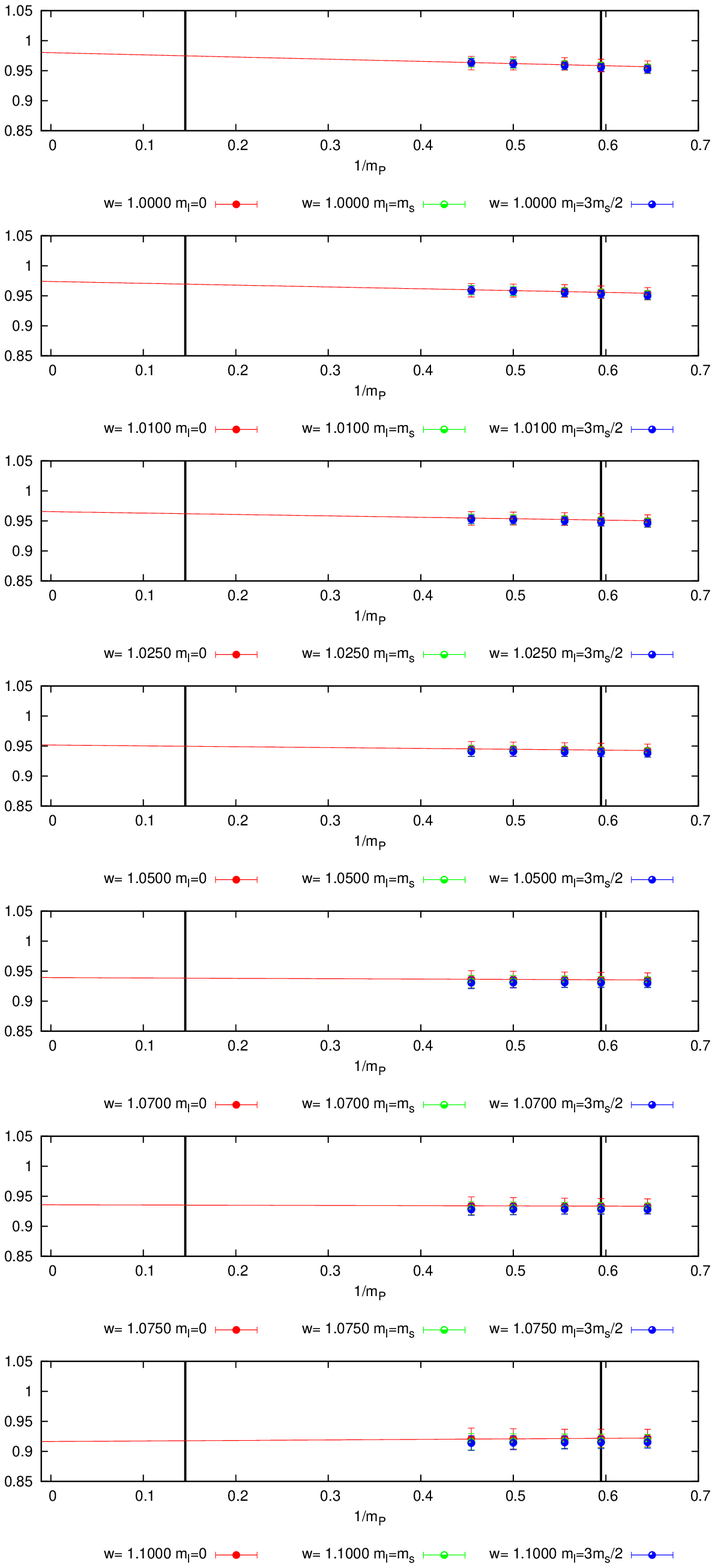}
\caption{\label{fig:references2} 
Continuum step scaling function $\sigma^{P\rightarrow D^\star}(w;L_1,L_2)$ as a function
of the inverse RGI heavy quark mass of the pseudoscalar state, $1/m_P$,
for the different values of $w$. The vertical black lines represent the physical values
of the charm and bottom quark masses. Symbols of different colors correspond to different
values of the light quark mass.}
\end{figure}

The last step of the finite size scaling procedure has been performed
by calculating the denominator of eq.~(\ref{eq:sigma2}) at three different lattice spacings,
corresponding to the entries $L_1a$, $L_1b$ and $L_1c$ of TABLE~\ref{tab:sims}, and
the numerator with the same lattice spacings but with twice the number of lattice points
along the spatial directions and three times the number of points along the time direction, 
entries $L_2A$, $L_2B$ and $L_2C$ of TABLE~\ref{tab:sims}. The choice of $T=3L/2$ in the
numerator of the last step is motivated by the need of reaching a time
extent in physical units that justifies the single state dominance hypothesis. This can also
be compared with the different sequence of steps, carried out in our previous 
papers~\cite{de Divitiis:2007ui,de Divitiis:2007uk}, reaching the same final time extent.

In FIG.~\ref{fig:continuums2} we show the continuum extrapolation of 
$\sigma^{P\rightarrow D^\star}(w;L_1,L_2)$ for the seven different values of $w$ that
have been simulated. The pseudoscalar meson state correspond to a heavy quark having a mass 
of about a quarter of the physical value of the $b$-quark
mass ($m_P\simeq m_b/4$), the heaviest mass simulated on these volumes; 
similar plots could have been shown for the other simulated heavy quark masses. 
In this step we have used a single definition of RGI quark masses and precisely
the one of eq.~(\ref{eq:rgi1}).

In FIG.~\ref{fig:references2} we show the extrapolations of the continuum step scaling function
$\sigma^{P\rightarrow D^\star}(w;L_1,L_2)$ to the physical point. 
The RGI heavy quark mass of the vector state has been held fixed to its physical value $m_c$.
Also in this second step the dependence of the finite volume effects upon the heavy
quark mass of the pseudoscalar state is very mild. The fact that we measure larger 
finite volume effects with respect to the ones obtained on the previous step 
can be explained by assuming that these are mainly due to the contribution of excited 
states to the correlation functions entering the definition of the lattice matrix elements
rather than to the matrix elements themselves. Indeed, in passing from the denominator 
to the numerator of eq.~(\ref{eq:sigma2}) the time extent of the final volume
is enlarged of a factor three, $T_2\simeq 2.4$~fm. The validity of this hypothesis is supported
by the results of a series of simulations of the volume $L_2^3\times L_2$ not shown in this paper. 

It has not been possible to simulate light quarks at vanishing mass on the final volume and
we have computed the step scaling function $\sigma^{P\rightarrow D^\star}(w;L_1,L_2)$ at
three different values of $m_l$ ranging from about $3m_s/2$ to about $m_s$, the physical value
of the strange quark mass. 
Symbols of different colors in the
plots of FIG.~\ref{fig:references2} show the step scaling functions for different
values of $m_l$. We do not see any appreciable dependence of finite volume effects upon
$m_l$ within the statistical errors and obtain results in the
chiral limit by a linear fit. The slopes of these fits are at least an order of magnitude
smaller than their error. 

\section{comparison with experiment}\label{sec:experiment}

\begin{table}[t]
\begin{ruledtabular}
\begin{tabular}{ccccc}
$w$ & $F^{B\rightarrow D^\star}(w)$ & $\frac{F^{B\rightarrow D^\star}(w)}{G^{B\rightarrow D}(w)}$ & $N_f$ & reference\\
\hline
\\[-5pt]
1.000 & 0.917(08)(05) & 0.878(10)(04) & 0   & this work \\
1.010 & 0.913(09)(05) & 0.883(09)(04) & 0   & this work \\
1.025 & 0.905(10)(05) & 0.891(09)(04) & 0   & this work \\
1.050 & 0.892(13)(04) & 0.905(10)(04) & 0   & this work \\
1.070 & 0.880(17)(04) & 0.914(12)(05) & 0   & this work \\
1.075 & 0.877(18)(04) & 0.916(12)(05) & 0   & this work \\
1.100 & 0.861(23)(04) & 0.923(16)(05) & 0   & this work \\
\\
\hline
1.00 & 0.913(20)(16) & & 0   &  \cite{Hashimoto:2001nb} \\
1.00 & 0.924(12)(19) & & 2+1 &  \cite{Laiho:2007pn}\\
\end{tabular}
\end{ruledtabular}
\caption{\label{tab:fw}
Final results in the continuum and
infinite volume limits. As a comparison we quote also the results of
previous lattice calculations of $F^{B\rightarrow D^\star}(w)$ by 
the Fermilab collaboration.}
\end{table}

Our final results for $F^{B\rightarrow D^\star}(w)$
are obtained by multiplying the small volume numbers with
the two step scaling functions according to eq.~(\ref{eq:infresultformula})
and are shown in TABLE~\ref{tab:fw}. 
The first error is statistical while the second
comes from the uncertainties associated to the renormalization factors on the small
volume (see discussion in section~\ref{tab:fw}). 
As a comparison we quote in table TABLE~\ref{tab:fw} the results of previous lattice calculations obtained by the Fermilab lattice collaboration~\cite{Laiho:2007pn,Hashimoto:2001nb} at zero recoil.

\begin{figure}[t]
\includegraphics[width=0.45\textwidth]{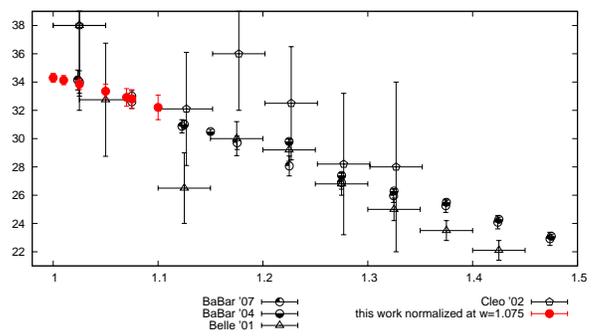}
\caption{\label{fig:vcb} 
Comparison of $\vert V_{cb}\times 10^3\vert F^{B\rightarrow D^\star}(w)$ obtained 
in this work with experimental data. 
Lattice data carry only statistical errors and are normalized at $w=1.075$.}
\end{figure}
\begin{figure}[t]
\includegraphics[width=0.45\textwidth]{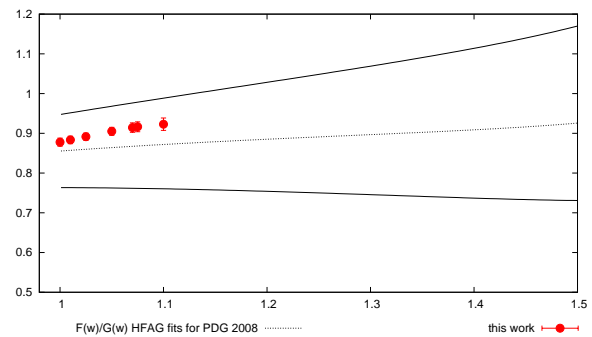}
\caption{\label{fig:ratio} 
Comparison of the ratio $F^{B\rightarrow D^\star}(w)/G^{B\rightarrow D}(w)$ obtained 
in this work with the fits to experimental data performed by the
HFAG for the 2008 edition of the PDG. Lattice data carry only statistical errors.}
\end{figure}

FIG.~\ref{fig:vcb} shows the comparison of our lattice data with some
experimental determinations of $\vert V_{cb}\times 10^3\vert F^{B\rightarrow D^\star}(w)$ 
\cite{Abe:2001cs,Briere:2002ew,Aubert:2004bw,Aubert:2007rs}: the functional dependence
of experimental data upon $w$ is reproduced by lattice data within statistical errors
also in the quenched approximation. The comparison is made by matching lattice data with 
the experimental ones from ref.~\cite{Aubert:2007rs} at $w=1.075$ and by obtaining
\begin{eqnarray}
\vert V_{cb} \vert = 3.74(8)(5)\times 10^{-2}
\end{eqnarray}
where the first error is from theory while the second from experiment.
This number is in good agreement with our previous determination of the CKM
matrix element, $V_{cb}=3.84(9)(42)\times 10^{-2}$, performed
in ref.~\cite{de Divitiis:2007ui} by matching $G^{B\rightarrow D}(w)$, 
the form factor entering the decay rate of the process $B\rightarrow D\ell\nu$, 
with experimental data at $w=1.2$. 
In both the cases $V_{cb}$ has been extracted at non vanishing momentum transfer
and the corresponding uncertainties are not correlated with the errors on the parameters
of the fits needed to extrapolate measured decay rates to zero recoil.  

A more direct comparison between lattice and experiment can be obtained by introducing
the ratio $F^{B\rightarrow D^\star}(w)/G^{B\rightarrow D}(w)$ that does not depend upon $V_{cb}$. 
In order to apply a jackknife procedure directly to the ratio of the two form
factors, we have computed $G^{B\rightarrow D}(w)$ on the same gauge ensambles
of $F^{B\rightarrow D^\star}(w)$ by following the finite size scaling recursion described in 
section~\ref{sec:ssm} and by using the definition of $G^{B\rightarrow D}(w)$ discussed
in refs.~\cite{de Divitiis:2007ui,de Divitiis:2007uk}. Our results
are given in TABLE~\ref{tab:fw} and compared with experiment in FIG.~\ref{fig:ratio}. 
The black curves have been drawn by using the 
two independent fits\footnote{A direct evaluation from
the experimental collaborations of the ratio of decay rates would likely lead to a
more stringent test.} of the experimental data for $F^{B\rightarrow D^\star}(w)$ and 
for $G^{B\rightarrow D}(w)$ performed by the Heavy Flavour Averaging Group~\cite{Barberio:2007cr}
by using the parametrizations of ref.~\cite{Caprini:1997mu}.
Our results, red points in the figure, fall inside the
allowed experimental region, i.e. between the two solid black curves.
Theoretical errors are smaller than experimental ones that in turn are
dominated by the uncertainties on $G^{B\rightarrow D}(w)$.
Also from this parameter-free comparison with experiments we obtain a strong
indication for the validity of the quenched approximation for these observables.
Together with the evidence of mild volume effects (if we exclude those coming from excited
states contributions to correlation functions) these results may call for a basic
short distance nature of the form factors. This could justify why main unquenching effects
are reabsorbed by the renormalization procedure of the quenched calculation that
fixes indirectly the renormalized coupling constant from physical quantities.

\begin{acknowledgments}
We warmly thank R. Faccini and F. Cossutti for providing us the data 
of figure 6 of ref.~\cite{Aubert:2007rs}.
The simulations required to carry on this project 
have been performed on the INFN apeNEXT machines at Rome "La Sapienza".
We thank A.~Lonardo, D.~Rossetti and P.~Vicini for technical advice. 
\end{acknowledgments}


\end{document}